\documentclass[12pt]{article}

\usepackage[margin=1in]{geometry}
\usepackage{amsmath,amssymb}
\usepackage{booktabs}
\usepackage{array}
\usepackage{graphicx}
\usepackage[hidelinks]{hyperref}
\usepackage{natbib}
\usepackage{setspace}
\usepackage{enumitem}
\usepackage[labelfont=bf]{caption}
\usepackage[T1]{fontenc}
\usepackage{microtype}
\usepackage{xspace}
\usepackage{threeparttable}

\newcommand{\mlmgof}{\texttt{mlm\_gof}\xspace}
\newcommand{\melogit}{\texttt{melogit}\xspace}
\newcommand{\nmin}{n_{\min}}

\title{\textbf{A Goodness-of-Fit Test for Mixed-Effects\\ 
Logistic Regression}}

\author{Ariel Linden\\[4pt]
  University of California, San Francisco\\
  Department of Medicine\\
  Division of Clinical Informatics \& Digital Transformation (DoC-IT)\\
  \texttt{ariel.linden@ucsf.edu}}

\date{}

\begin{document}
\maketitle
\thispagestyle{empty}

\begin{abstract}
\noindent
Mixed-effects logistic regression is widely used for analyzing binary outcomes in hierarchical data, yet formal goodness-of-fit tests remain limited to random intercept models and do not address sparse cluster settings. We extend a grouping-based Wald test to mixed-effects logistic models with random slopes. The procedure groups observations by their predicted probabilities within clusters, adds pooled group indicators to the model, and tests their joint significance using a Wald statistic. To accommodate small clusters, we introduce a data-driven rule for selecting the number of groups, $G = \min(10, \nmin)$, where $\nmin$ is the smallest cluster size. This prevents estimation failure when clusters contain fewer than ten observations. Simulation studies across 24 null scenarios show that the test maintains nominal Type~I error in three-level random slope models, including at smaller sample sizes than previously studied. The test demonstrates increasing power to detect fixed-effects misspecification: power against omitted nonlinearity increased from 0.07 to 1.00 across effect sizes, and power against omitted interactions reached 0.87. As expected, the test has no power to detect an omitted clustering level, reflecting its focus on residual patterns in predicted probabilities. In sparse balanced designs, fixing $G=10$ led to complete test failure, whereas the data-driven rule performed reliably. The method is implemented in the
community-contributed Stata program \mlmgof after fitting a mixed-effects logistic regression model.

\end{abstract}

\newpage\setcounter{page}{1}

\section{Introduction}\label{sec:intro}

Multilevel data structures are common in health services, epidemiologic, and social science research. Patients are nested within doctors, hospitals within regions, and repeated measurements within individuals. When the outcome is binary (e.g., mortality, disease onset, or treatment success) the appropriate analytic tool is the mixed-effects logistic regression model, which accounts for clustering and partitions variation across
levels. The terms \emph{mixed-effects}, \emph{multilevel}, and \emph{hierarchical} are used interchangeably in the literature; here we use \emph{mixed-effects}. Similarly, we use \emph{random slopes} to refer to models in which regression coefficients vary across clusters. The use of these models has increased substantially in recent decades across
medicine, public health, and the social sciences.\textsuperscript{1}

A fitted model is only useful to the extent that it adequately describes the data. Before drawing inferences about covariate effects or variance components, it is essential to assess model fit. For single-level logistic regression, several goodness-of-fit tests are well established. The Hosmer--Lemeshow test\textsuperscript{2,3} groups observations by predicted probabilities and compares observed and expected frequencies. Lipsitz et al.\textsuperscript{4} extended this approach by adding group indicator variables to the model and testing their joint significance via a Wald statistic, making the method adaptable to correlated data. Further extensions have addressed survey-weighted models,\textsuperscript{5} complex
sampling designs,\textsuperscript{6} and multinomial outcomes.\textsuperscript{7}

In contrast, goodness-of-fit testing for mixed-effects logistic regression has been limited. Cool et al.\textsuperscript{8} highlighted the absence of suitable methods for higher-level multilevel binary models. Sturdivant and Hosmer\textsuperscript{9} proposed residual-based statistics but did not evaluate power. Perera et al.\textsuperscript{10} developed the first formal test for two-level models by adapting grouping-based methods to
clustered data. Fernando and Sooriyarachchi\textsuperscript{11} extended this framework to three-level models using limited-information approaches. Although these contributions represent important advances, they are restricted to random intercept models and do not address the more general case in which both intercepts and slopes vary randomly across
clusters, which is the appropriate specification whenever the effect of a covariate may differ across clusters.

A second limitation of existing tests is the fixed use of ten groups in the grouping procedure, following Hosmer and Lemeshow.\textsuperscript{2} In mixed-effects settings, clusters may contain fewer than ten observations, making this choice infeasible: empty cells arise, indicator variables become degenerate, and the augmented model cannot be estimated. Existing methods do not address this issue or provide guidance for sparse
cluster settings.

This paper makes two contributions. First, we extend the grouping-based Wald test framework\textsuperscript{10,11} to mixed-effects logistic regression models with random slopes. Second, we introduce a data-driven rule for selecting the number of groups, $G = \min(10, \nmin)$, where $\nmin$ is the minimum cluster size. Simulation results show that this rule is necessary to avoid test failure when clusters are small. We evaluate the
proposed method using an extensive simulation study covering Type~I error, power under three types of misspecification, and a direct comparison of the data-driven grouping rule with the conventional choice of $G=10$. The method is implemented in the community-contributed Stata program \mlmgof.\textsuperscript{12}

The remainder of the paper is organized as follows. Section~\ref{sec:methods} describes the model, test procedure, and simulation design. Section~\ref{sec:results} reports simulation results. Section~\ref{sec:example} presents an applied example. 
Section~\ref{sec:discussion} discusses results and Section~\ref{sec:conclusions} concludes.

\section{Methods}\label{sec:methods}

\subsection{The Mixed-Effects Logistic Regression Model with Random Slopes}
\label{sec:model}

Let $y_{ij}$ denote a binary outcome for observation $i$ in cluster $j$
($i = 1,\ldots,n_j$; $j = 1,\ldots,J$). A two-level mixed-effects logistic regression model with a random intercept and a random slope on covariate $x_{2ij}$ is 
\begin{equation}
  \operatorname{logit}(p_{ij}) = \beta_0 + \beta_1 x_{1ij} + \beta_2 x_{2ij}
    + u_{0j} + u_{1j}\,x_{2ij},
  \label{eq:baseline}
\end{equation}
where $p_{ij} = \Pr(y_{ij}=1\mid\mathbf{u}_j)$, $\beta_0, \beta_1, \beta_2$ are fixed effects, and $(u_{0j}, u_{1j})^\top \sim N(\mathbf{0}, \boldsymbol{\Omega})$ are cluster-specific random effects. The random intercept $u_{0j}$ captures between-cluster variation in baseline log-odds, while the random slope $u_{1j}$ allows the effect of $x_2$ to vary across clusters. Conditional predicted probabilities are obtained as
\begin{equation}
  \hat{p}_{ij} = \operatorname{logit}^{-1}\!\bigl(
    \hat{\beta}_0 + \hat{\beta}_1 x_{1ij} + \hat{\beta}_2 x_{2ij}
    + \tilde{u}_{0j} + \tilde{u}_{1j}\,x_{2ij}\bigr),
  \label{eq:phat}
\end{equation}
where $\tilde{u}_{0j}$ and $\tilde{u}_{1j}$ are empirical Bayes estimates. A three-level extension adds random effects at an intermediate level (e.g., subjects within families). This formulation generalizes prior work, which considered random intercept models only.\textsuperscript{10,11}

\subsection{Derivation of the Goodness-of-Fit Test Statistic}
\label{sec:teststat}

The test follows the model-based approach of Lipsitz et al.,\textsuperscript{4} extended to clustered data by Perera et al.\textsuperscript{10} for the two-level case and by Fernando and Sooriyarachchi\textsuperscript{11} for the three-level case. The null
hypothesis is that the fitted model adequately describes the data. Under the null, the predicted probabilities $\hat{p}_{ij}$ should capture the observed outcomes without systematic residual structure.

Let $G$ denote the number of groups formed from the distribution of $\hat{p}_{ij}$ within each cluster. Define $I^g_{ij} = 1$ if observation $i$ in cluster $j$ belongs to group $g$ ($g = 2,\ldots,G$). The augmented model is
\begin{equation}
  \operatorname{logit}(p_{ij}) = \beta_0 + \beta_1 x_{1ij} + \beta_2 x_{2ij}
    + \gamma_2 I^2_{ij} + \cdots + \gamma_G I^G_{ij}
    + u_{0j} + u_{1j}\,x_{2ij}.
  \label{eq:augmented}
\end{equation}
The null hypothesis is $H_0\colon \gamma_2 = \cdots = \gamma_G = 0$. The joint Wald statistic is
\begin{equation}
  W = \hat{\boldsymbol{\gamma}}^\top
    \bigl[\widehat{\operatorname{Var}}(\hat{\boldsymbol{\gamma}})\bigr]^{-1}
    \hat{\boldsymbol{\gamma}},
  \label{eq:wald}
\end{equation}
which is approximately $\chi^2$ with $G-1$ degrees of freedom.\textsuperscript{4} Large values of $W$ indicate lack of fit.

\subsection{The Step-by-Step Procedure Implemented in \mlmgof}
\label{sec:procedure}

\begin{enumerate}[leftmargin=*, label=\textbf{Step \arabic*.}, itemsep=4pt]

\item \textbf{Fit the baseline model.} Estimate the mixed-effects logistic model using maximum likelihood and obtain predicted probabilities $\hat{p}_{ij}$.

\item \textbf{Select the number of groups.} Set $G = \min(10, \nmin)$, where $\nmin$ is the minimum number of observations in any level-2 cluster.

\item \textbf{Group observations.} Within each level-2 cluster, sort $\hat{p}_{ij}$ and assign observations to $G$ approximately equal-sized groups using the asymptotic ranking approach of Rosner et al.,\textsuperscript{13} which preserves the overall ranking across
  clusters while enabling within-cluster grouping.

\item \textbf{Create indicators.} For each observation, define $G-1$ binary indicator variables $I^g_{ij} = 1$ if observation $i$ in cluster $j$ belongs to group $g$ ($g = 2, \ldots, G$), with group 1 as the reference. Indicators for the same group $g$ across clusters are pooled into a single variable for the full dataset.\textsuperscript{10}

\item \textbf{Refit the model.} Estimate the augmented model~(\ref{eq:augmented}) including the indicators and the original random-effects structure.

\item \textbf{Compute the test statistic.} Compute $W$ from equation~(\ref{eq:wald}) and compare to a $\chi^2_{G-1}$ distribution to obtain the $p$-value.

\item \textbf{Report results.} Report $W$, degrees of freedom, $p$-value, and the value of $G$ used.

\end{enumerate}

\noindent This implementation differs from prior work\textsuperscript{10,11} in that estimation uses full maximum likelihood and the original random-effects structure is retained in the augmented model.

\subsection{The Data-Driven Group Selection Rule}
\label{sec:grouprule}

The conventional choice $G=10$ is not always feasible in mixed-effects settings where clusters may be small. When $\nmin < 10$, dividing clusters into ten groups produces empty cells and prevents estimation. We therefore use $G = \min(10, \nmin)$, which ensures that each group contains at least one observation per cluster. When clusters are sufficiently
large, this reduces to the standard choice. Simulation results
(Section~\ref{sec:groupsensitivity}) show that this rule is necessary for valid inference in sparse cluster settings.

\subsection{Simulation Study Design}
\label{sec:simdesign}

We conducted three simulation studies to evaluate the proposed test. The data-generating process was common to all components; design parameters are summarized in Tables~\ref{tab:sim1design}--\ref{tab:sim3design}. The data-generating process and null model are three-levels throughout, as this is the general case: a two-level model is a special case of a three-level model in which the intermediate level has a single unit per cluster, so adequate performance under three-level specifications implies adequate
performance for two-level models as well. 1000 replications were generated for each of the following 44 scenarios (total of 44,000 replications).

\medskip
\noindent\textbf{Data-generating process.} The true model is
\begin{equation}
  \operatorname{logit}(p_{ijk}) = \beta_0 + \beta_1 x_{1ijk} + \beta_2 x_{2ijk}
    + v_j + u_{kj} + w_{kj}\,x_{2ijk},
  \label{eq:dgp}
\end{equation}
where $x_{1ijk} \sim U(-3,3)$, $x_{2ijk} \sim \mathrm{Bernoulli}(0.5)$,
$v_j \sim N(0,\sigma^2_v)$ is the family-level random intercept,
$u_{kj} \sim N(0,\sigma^2_u)$ is the subject-level random intercept, and
$w_{kj} \sim N(0,\sigma^2_w)$ is the subject-level random slope on $x_2$, all mutually
independent. Fixed parameters: $\beta_0 = -1.0$, $\beta_1 = 0.5$, $\beta_2 = 0.3$.
ICC $= \sigma^2/(\sigma^2 + \pi^2/3)$.

\medskip
\noindent\textbf{Part~1: Type I error.} Data are generated from the correctly specified model and fitted by Stata's \melogit command. Factors varied: $J \in \{15,30,50\}$, $K \in \{5,10\}$, $n=20$, ICC $\in \{0.10, 0.30\}$, accounting for 24 scenarios. See Table~\ref{tab:sim1design}.

\medskip
\noindent\textbf{Part~2: Power.} Fixed conditions: $J=30$, $K=5$, $n=20$,
ICC $= 0.20$. Three misspecification types: (a)~omitted quadratic term
($\beta_3 x_1^2$, $\beta_3 \in \{0.02, 0.05, 0.10, 0.15\}$); (b)~omitted interaction ($\beta_3(x_1 \times x_2)$, $\beta_3 \in \{0.3, 0.6, 0.9\}$); (c)~omitted level (true three-level model fitted as two-level, $\sigma_{\text{extra}} \in \{0.5, 1.0, 1.5\}$), for a total of 10 scenarios. See Table~\ref{tab:sim2design}.

\medskip
\noindent\textbf{Part~3: Group selection sensitivity.} $G = \min(10,\nmin)$ vs.\ $G=10$
at $n_{\text{small}} \in \{3,5,6,8,10\}$. Unbalanced design: 10 small clusters ($n_{\text{small}}$ obs) and 40 large clusters ($n=20$). Balanced design: all 50 clusters have $n_{\text{small}}$ observations. Both designs together account for a total of 10 scenarios. See Table~\ref{tab:sim3design}.

\medskip
\noindent This design differs from prior work\textsuperscript{10,11} in three respects: estimation uses full maximal likelihood rather than penalized or marginal quasi-likelihood methods (PQL/MQL); the addition of random slopes in the model; and smaller sample sizes ($N = 1{,}500$--$7{,}500$) than the minimum $N=4{,}500$ used by Fernando and Sooriyarachchi\textsuperscript{11}, allowing for validation at more modest sample sizes than previously studied. All analyses were conducted using Stata version~19 (StataCorp LLC, College Station, TX).

\section{Results}\label{sec:results}

\subsection{Type I Error (Part~1)}\label{sec:typeIerror}

\begin{figure}[t]
  \centering
  \includegraphics[width=0.88\textwidth]{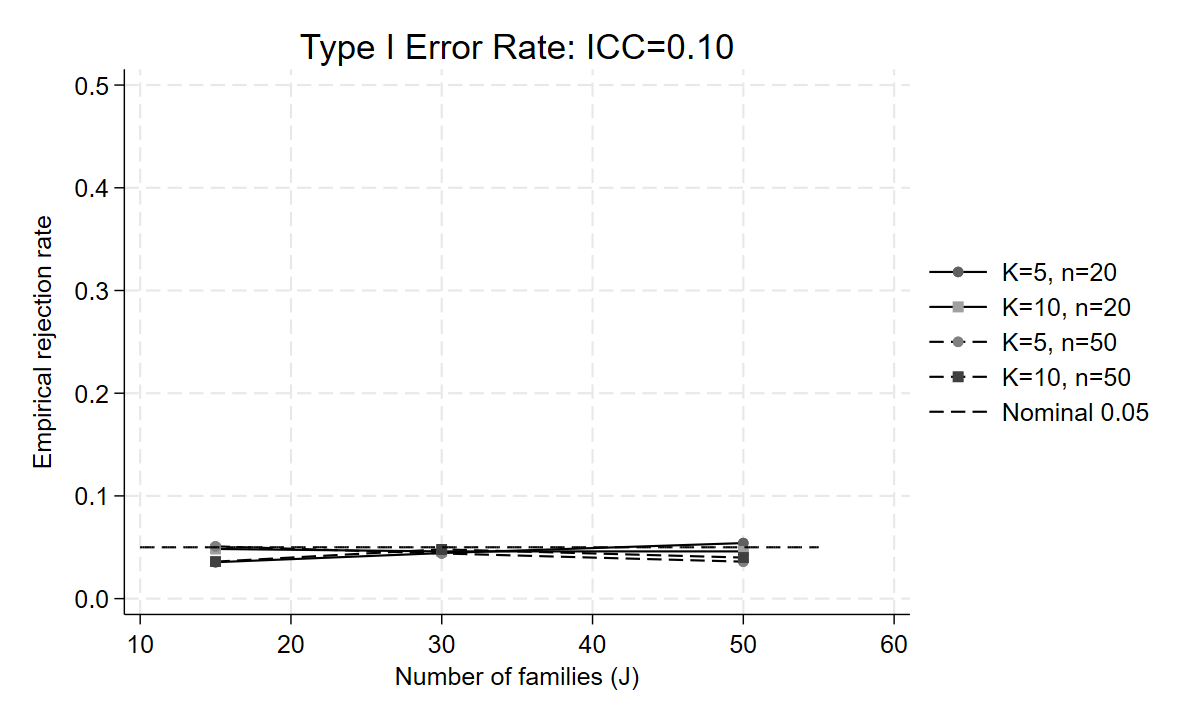}
  \caption{Empirical Type~I error rate by number of families ($J$), subjects per family
    ($K$), and observations per subject ($n$). ICC $= 0.10$. Dashed line = nominal 0.05.
    All values within Monte Carlo bounds $[0.036, 0.064]$.}
  \label{fig:typeI_010}
\end{figure}

\begin{figure}[t]
  \centering
  \includegraphics[width=0.88\textwidth]{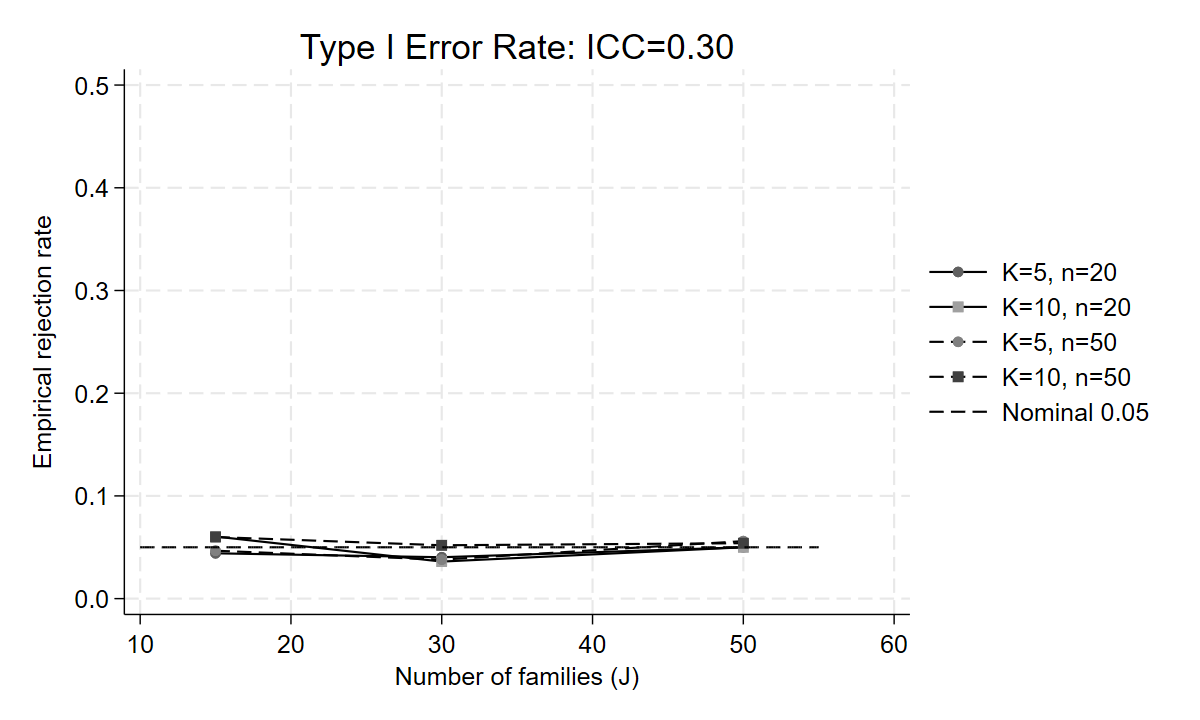}
  \caption{Empirical Type~I error rate by number of families ($J$), subjects per family
    ($K$), and observations per subject ($n$). ICC $= 0.30$. Dashed line = nominal 0.05.
    All values within Monte Carlo bounds $[0.036, 0.064]$.}
  \label{fig:typeI_030}
\end{figure}

Figures~\ref{fig:typeI_010} and~\ref{fig:typeI_030} present empirical rejection rates of \mlmgof at $\alpha=0.05$ across all 24 null scenarios. Rejection rates ranged from 0.035 to 0.060, with all values within the Monte Carlo bounds of $[0.036, 0.064].$\textsuperscript{14} No systematic pattern of inflation or conservatism was observed. Performance remained stable at the smallest design ($J=15$, $K=5$, $n=20$; $N=1{,}500$), indicating that the test maintains nominal Type~I error even at sample sizes smaller than those considered in prior studies.\textsuperscript{10,11}

\subsection{Power (Part~2)}\label{sec:power}

Table~\ref{tab:power} summarizes empirical power across the three misspecification
scenarios.

\medskip\noindent\textbf{Omitted quadratic term.} Power was low at the smallest effect ($\beta_3 = 0.02$, power $= 0.074$), consistent with negligible misspecification, and increased rapidly, exceeding 0.90 at $\beta_3 = 0.10$, indicating strong sensitivity to nonlinear functional form misspecification.

\medskip\noindent\textbf{Omitted interaction.} Power increased more gradually, reaching 0.54 at $\beta_3 = 0.6$ and 0.87 at $\beta_3 = 0.9$, consistent with moderate-to-strong sensitivity at larger effect sizes.

\medskip\noindent\textbf{Omitted clustering level.} Power remained low
across all values of $\sigma_{\text{extra}}$, indicating that the test does not detect misspecification arising from omission of a clustering level.

\subsection{Group Selection Sensitivity (Part~3)}\label{sec:groupsensitivity}

Results are shown in Table~\ref{tab:groupsensitivity}.

\medskip\noindent\textbf{Unbalanced design.} Both the data-driven rule and $G=10$ maintained acceptable Type~I error across all values of $n_{\text{small}}$. Differences were minimal, reflecting the dominance of larger clusters in the test statistic.

\medskip\noindent\textbf{Balanced design.} The data-driven rule produced valid results across all scenarios within Monte Carlo bounds. In contrast, fixing $G=10$ failed in all replications when $n_{\text{small}} < 10$, yielding no valid test results. When $n_{\text{small}} = 10$, both approaches produced identical results.

\section{Applied Example}\label{sec:example}

\subsection{Study Context and Data}

We illustrate the implementation of \mlmgof using an artificial dataset reflecting a realistic disease management study. Disease management programs target individuals with chronic conditions or elevated risk, through structured behavioral and clinical interventions.\textsuperscript{15} Pre-diabetes is a common target given the effectiveness of lifestyle interventions in preventing progression to type~2 diabetes.\textsuperscript{16} 
The dataset represents a family-based intervention study in which patients with pre-diabetes are enrolled along with family members, yielding a three-level structure: repeated clinic visits (level 1) nested within subjects (level 2) nested within families (level 3). The binary outcome is whether pre-diabetes was reversed at each follow-up visit. The dataset includes approximately 900 observations across 30 families, with 5–7 subjects per family and 5 visits per subject. Covariates include intervention assignment (1 = intervention, 0 = control), body mass index centered at 30 kg/m² (bmi\_c), and visit number (1–5). The data-generating model includes a family-level random intercept, a family-level random slope on visit, and a subject-level random intercept.

\subsection{Model Fitting and Goodness-of-Fit Assessment}

We fit the model
\begin{equation}
  \operatorname{logit}(p_{ijk}) = \beta_0 + \beta_1\,\text{intervention}
    + \beta_2\,\text{bmi\_c} + \beta_3\,\text{visit}
    + v_j + w_j\,\text{visit} + u_{kj},
  \label{eq:example}
\end{equation}
where $v_j$ is the family-level random intercept, $w_j$ is the family-level random slope for visit, and $u_{kj}$ is the subject-level random intercept. Table~\ref{tab:example} reports the results. Because each subject contributes five observations, $\nmin = 5$ and the data-driven rule sets $G=5$.

The goodness-of-fit test yields $W = 2.92$ with $df=4$ ($p=0.404$), providing no evidence against adequate fit, as expected since the fitted model corresponds to the data-generating process. Using $G=10$ would lead to estimation failure in this setting
(Section~\ref{sec:groupsensitivity}).

\section{Discussion}\label{sec:discussion}

We developed and validated a goodness-of-fit test for mixed-effects logistic regression models. The test extends the grouping-based Wald framework of Perera et al.\textsuperscript{10} and Fernando and Sooriyarachchi\textsuperscript{11} to accommodate both random slopes and sparse cluster settings. It is implemented in the Stata program \mlmgof, allowing direct application after fitting a mixed-effects logistic regression model.

The simulation results yield three principal findings. First, the test maintains nominal Type~I error across a range of three-level random slope specifications. This holds at the smallest design ($N=1{,}500$), considerably smaller than the minimum $N=4{,}500$ at which
Fernando and Sooriyarachchi\textsuperscript{11} reported instability, and smaller than the settings where Perera et al.\textsuperscript{10} observed distortion (rejection rate 0.013). Second, the test has good power against fixed-effects misspecification, including omitted nonlinear terms and omitted interactions. Both prior studies evaluated power using a single $\log X^2$ alternative that yields near-perfect power in large
samples but does not characterize the sensitivity profile across effect sizes; the graduated grids used here are more informative. Third, the test has no power to detect omission of a clustering level, regardless of the magnitude of the omitted variance component. This finding reflects a structural limitation of the method rather than a deficiency of the simulation design. The grouping procedure detects systematic patterns in the conditional predicted probabilities within clusters; omission of a random intercept does not induce such patterns. As a result, goodness-of-fit tests based on grouping cannot detect this form of misspecification. Researchers concerned about the adequacy of the random-effects structure should instead rely on likelihood-based model comparison or information criteria.

The data-driven group selection rule $G = \min(10, \nmin)$ is the second major contribution of this work. The simulation results demonstrate that this rule is necessary, not merely convenient. In balanced cluster designs with n\_min<10, fixing $G=10$ led to complete failure of the test in every replication, producing no valid results. In contrast, the data-driven rule yielded valid inference across all scenarios while recovering the conventional choice when clusters were sufficiently large. Neither Perera et al.\textsuperscript{10} nor Fernando and Sooriyarachchi\textsuperscript{11} address this issue, instead they adopt the standard choice of $G=10$ from Hosmer and Lemeshow.\textsuperscript{2} The present findings show that this convention does not generalize to mixed-effects settings with small clusters.

The improved performance of the proposed test relative to prior work likely reflects two factors. First, the use of full maximum likelihood estimation via adaptive Gauss–Hermite quadrature yields more reliable parameter estimates than the penalized or marginal quasi-likelihood methods (PQL/MQL) used in earlier studies, which are known to produce biased estimates in binary mixed-effects models, particularly in small samples or with high intraclass correlation.\textsuperscript{17,18} Second, the extension to random slope models allows the test to be applied in settings where covariate effects vary across clusters, which is often the appropriate specification in practice.

Several limitations should be noted. The Wald statistic relies on asymptotic approximations in the number of clusters; when the number of clusters is small (e.g., $J < 15$), the $\chi^2$ approximation may be unreliable, and bootstrap or permutation-based approaches may be preferable. The test is also sensitive to the choice of the number of groups $G$, as both the degrees of freedom and the grouping structure influence power; sensitivity analysis across alternative values of $G$ is straightforward using the \texttt{groups()} option in \mlmgof. Finally, the method is developed for binary outcomes; extensions to ordinal, count, or time-to-event outcomes in mixed-effects models remain topics for future research.

From an applied perspective, \mlmgof addresses a gap in the diagnostic toolkit for mixed-effects logistic regression. A non-significant result indicates that grouping of predicted probabilities does not reveal systematic lack of fit, providing a useful complement to other diagnostics such as residual analysis and likelihood-based model comparison. In combination, these tools support a more comprehensive assessment of model adequacy.

More broadly, the evaluation of goodness-of-fit is an essential component of rigorous statistical analysis. As Linden and Roberts\textsuperscript{19} emphasize, the validity of conclusions drawn from fitted models depends on demonstrating that those models adequately represent the data. In mixed-effects logistic regression, where model structures are more complex and assumptions less transparent than in single-level analyses, the availability of a formal goodness-of-fit test is particularly valuable. The proposed method facilitates routine assessment and reporting of model fit in applied research using mixed-effects logistic models.

\section{Conclusions}\label{sec:conclusions}

We have presented a goodness-of-fit test for mixed-effects logistic regression models that include random slopes, implemented in the Stata program \mlmgof. The test extends the grouping-based Wald framework designed only for models with random intercepts and additionally introduces a data-driven group selection rule that prevents failure in sparse cluster contexts. It is validated across a broad range of multilevel designs. Simulation results demonstrate excellent Type I error control, good power to detect fixed-effects misspecification, and the necessity of the data-driven rule when cluster sizes fall below ten. The test is sensitive to functional form misspecification but not to omitted random effects, a distinction that should guide its use alongside likelihood-based model comparison tools. Together, these properties make \mlmgof a practical and principled tool for assessing model adequacy in mixed-effects logistic regression.

\clearpage
\section*{References}

\begin{enumerate}[leftmargin=*, label={[\arabic*]}, itemsep=2pt]

\item Austin PC, Merlo J. Intermediate and advanced topics in multilevel logistic
  regression analysis. \textit{Stat Med}. 2017;36(20):3257--3277.

\item Hosmer DW, Lemeshow S. A goodness-of-fit test for the multiple logistic regression
  model. \textit{Commun Stat Theory Methods}. 1980;9(10):1043--1069.

\item Hosmer DW, Lemeshow S. \textit{Applied Logistic Regression}. 2nd ed. New York, NY:
  John Wiley \& Sons; 2000.

\item Lipsitz SR, Fitzmaurice GM, Molenberghs G. Goodness-of-fit tests for ordinal
  response regression models. \textit{J R Stat Soc Ser C Appl Stat}.
  1996;45(2):175--190.

\item Archer KJ, Lemeshow S. Goodness-of-fit test for a logistic regression model fitted
  using survey sample data. \textit{Stata J}. 2006;6(1):97--105.

\item Archer KJ, Lemeshow S, Hosmer DW. Goodness-of-fit tests for logistic regression
  models when data are collected using a complex sampling design. \textit{Comput Stat
  Data Anal}. 2007;51:4450--4464.

\item Fagerland MW, Hosmer DW, Bofin AM. Multinomial goodness-of-fit tests for logistic
  regression models. \textit{Stat Med}. 2008;27:4238--4253.

\item Cool G, Lebel A, Sadiq R, Rodriguez MJ. Modelling the regional variability of the
  probability of high trihalomethane occurrence in municipal drinking water.
  \textit{Environ Monit Assess}. 2015;187(12):746.

\item Sturdivant RX, Hosmer DW. Smoothed residual based goodness-of-fit statistics for
  logistic hierarchical regression models. \textit{Comput Stat Data Anal}.
  2007;51(8):3898--3912.

\item Perera AAPNM, Sooriyarachchi MR, Wickramasuriya SL. A goodness of fit test for the
  multilevel logistic model. \textit{Commun Stat Simul Comput}. 2016;45(2):643--659.

\item Fernando G, Sooriyarachchi R. The development of a goodness-of-fit test for high
  level binary multilevel models. \textit{Commun Stat Simul Comput}.
  2022;51(5):2710--2730.

\item Linden A. MLM\_GOF: Stata module for computing the goodness-of-fit test after
  mixed-effects logistic regression. Statistical Software Components S459670. Boston
  College Department of Economics; 2026.

\item Rosner B, Glynn RJ, Lee M-LT. Incorporation of clustering effects for the Wilcoxon
  rank sum test: a large-sample approach. \textit{Biometrics}. 2003;59(4):1089--1098.

\item Fleiss JL. \textit{Statistical Methods for Rates and Proportions}. 2nd ed. New York,
  NY: John Wiley \& Sons; 1981.

\item Linden A, Adler-Milstein J. Medicare disease management in policy context.
  \textit{Health Care Financ Rev}. 2008;29(3):1--11.

\item Biuso TJ, Butterworth S, Linden A. A conceptual framework for targeting prediabetes
  with lifestyle, clinical and behavioral management interventions. \textit{Dis Manag}.
  2007;10(1):6--15.

\item Goldstein H, Rasbash J. Improved approximations for multilevel models with binary
  responses. \textit{J R Stat Soc Ser A Stat Soc}. 1996;159(3):505--513.

\item Browne WJ, Draper D. A comparison of Bayesian and likelihood-based methods for
  fitting multilevel models. \textit{Bayesian Anal}. 2006;1(3):473--514.

\item Linden A, Roberts N. A user's guide to the disease management literature:
  recommendations for reporting and assessing program outcomes. \textit{Am J Manag Care}.
  2005;11(2):113--120.

\end{enumerate}

\clearpage


\begin{table}[ht]
\caption{Type I Error Simulation Design}
\label{tab:sim1design}
\centering
\begin{threeparttable}
\begin{tabular}{ll}
\toprule
Factor & Values \\
\midrule
Number of clusters ($J$)       & 15, 30, 50 \\
Subjects per cluster ($K$)     & 5, 10 \\
Observations per subject ($n$) & 20 \\
Intraclass correlation (ICC)   & 0.10, 0.30 \\
Total scenarios                & 24 \\
Replications per scenario      & 1000\tnote{*} \\
Significance level             & 0.05 \\
\bottomrule
\end{tabular}
\begin{tablenotes}\small
\item[*] Monte Carlo bounds for 1000 replications at $\alpha=0.05$: $[0.036, 0.064]$.
\end{tablenotes}
\end{threeparttable}
\end{table}

\begin{table}[ht]
\caption{Power Simulation Design}
\label{tab:sim2design}
\centering
\begin{threeparttable}
\begin{tabular}{p{3.2cm}p{3.5cm}p{3.2cm}p{1.8cm}p{3.5cm}}
\toprule
Misspecification & DGP & Fitted model & Parameter & Values \\
\midrule
Omitted quadratic  & Includes $\beta_3 x_1^2$           & Omits $\beta_3 x_1^2$           & $\beta_3$              & 0.02, 0.05, 0.10, 0.15 \\[4pt]
Omitted interaction & Includes $\beta_3(x_1{\times}x_2)$ & Omits $\beta_3(x_1{\times}x_2)$ & $\beta_3$              & 0.3, 0.6, 0.9 \\[4pt]
Omitted level      & 3-level model                       & 2-level model                   & $\sigma_{\text{extra}}$ & 0.5, 1.0, 1.5 \\
\bottomrule
\end{tabular}
\begin{tablenotes}\small
\item Fixed conditions: $J=30$, $K=5$, $n=20$, ICC$=0.20$; 1000 replications per scenario.
\end{tablenotes}
\end{threeparttable}
\end{table}

\begin{table}[ht]
\caption{Group Selection Simulation Design}
\label{tab:sim3design}
\centering
\begin{threeparttable}
\begin{tabular}{p{3cm}p{4cm}p{5.5cm}}
\toprule
\multicolumn{3}{l}{\textit{Cluster designs}} \\
\midrule
Design & Clusters & Cluster sizes \\
\midrule
Unbalanced & 10 small; 40 large & Small: $n_{\text{small}}\in\{3,5,6,8,10\}$; Large: $n=20$ \\
Balanced   & 50 clusters        & All: $n_{\text{small}}\in\{3,5,6,8,10\}$ \\
\midrule
\multicolumn{3}{l}{\textit{Evaluation settings}} \\
\midrule
\multicolumn{2}{l}{Grouping rules compared}   & $G=\min(10,\nmin)$ vs.\ $G=10$ \\
\multicolumn{2}{l}{Outcome measures}          & Type I error; failure rate \\
\multicolumn{2}{l}{Replications per scenario} & 1000 \\
\multicolumn{2}{l}{Failure definition}        & Missing $p$-value \\
\bottomrule
\end{tabular}
\end{threeparttable}
\end{table}

\begin{table}[ht]
\caption{Empirical power of the goodness-of-fit test against three types of model
  misspecification}
\label{tab:power}
\centering
\begin{threeparttable}
\begin{tabular}{llcc}
\toprule
Misspecification & Parameter & Effect size & Power \\
\midrule
Omitted quadratic & $\beta_3$ & 0.02 & 0.074 \\
                  &           & 0.05 & 0.278 \\
                  &           & 0.10 & 0.923 \\
                  &           & 0.15 & 0.998 \\
\addlinespace
Omitted interaction & $\beta_3$ & 0.3 & 0.143 \\
                    &           & 0.6 & 0.544 \\
                    &           & 0.9 & 0.874 \\
\addlinespace
Omitted level & $\sigma_{\text{extra}}$ & 0.5 & 0.043 \\
              &                        & 1.0 & 0.060 \\
              &                        & 1.5 & 0.044 \\
\bottomrule
\end{tabular}
\begin{tablenotes}\small
\item Fixed conditions: $J=30$, $K=5$, $n=20$, ICC$=0.20$. Power for omitted level
  scenarios is indistinguishable from the nominal $\alpha=0.05$.
\end{tablenotes}
\end{threeparttable}
\end{table}

\begin{table}[ht]
\caption{Type I error rates: data-driven $G=\min(10,\nmin)$ versus $G=10$, by minimum
  cluster cell size}
\label{tab:groupsensitivity}
\centering
\begin{threeparttable}
\begin{tabular}{cccccc}
\toprule
 & & \multicolumn{2}{c}{Unbalanced design} & \multicolumn{2}{c}{Balanced design} \\
\cmidrule(lr){3-4}\cmidrule(lr){5-6}
$\nmin$ & $G$ & Rej.\ default & Rej.\ $G=10$ & Rej.\ default & Rej.\ $G=10$ \\
\midrule
 3 & 3  & 0.040 & 0.034 & 0.028 & --- \\
 5 & 5  & 0.044 & 0.050 & 0.034 & --- \\
 6 & 6  & 0.062 & 0.044 & 0.050 & --- \\
 8 & 8  & 0.032 & 0.040 & 0.038 & --- \\
10 & 10 & 0.048 & 0.048 & 0.030 & 0.030 \\
\bottomrule
\end{tabular}
\begin{tablenotes}\small
\item Unbalanced: $J_{\text{small}}=10$ clusters with $\nmin$ obs and
  $J_{\text{large}}=40$ clusters with $n=20$. Balanced: all $J=50$ clusters have $\nmin$
  obs. ``---'' = $G=10$ produced no valid result for all replications. Monte Carlo bounds:
  $[0.036, 0.064]$.
\end{tablenotes}
\end{threeparttable}
\end{table}

\begin{table}[ht]
\caption{Mixed-effects logistic regression results: pre-diabetes reversal study ($N=905$)}
\label{tab:example}
\centering
\begin{threeparttable}
\begin{tabular}{lcccc}
\toprule
Parameter & OR & SE & 95\% CI & $p$ \\
\midrule
\multicolumn{5}{l}{\textit{Fixed effects}} \\
\quad Intervention & 5.786 & 1.293 & (3.734, 8.966) & $<$0.001 \\
\quad BMI centered (per unit above 30~kg/m\textsuperscript{2}) & 0.901 & 0.032 & (0.840, 0.967) & 0.004 \\
\quad Visit number & 1.313 & 0.136 & (1.071, 1.609) & 0.009 \\
\quad Intercept (baseline odds)\tnote{a} & 0.147 & 0.042 & (0.084, 0.256) & $<$0.001 \\
\midrule
\multicolumn{5}{l}{\textit{Random effects (SD)}} \\
\quad Family-level: intercept & 0.741 & 0.298 & (0.336, 1.631) & --- \\
\quad Family-level: slope (visit) & 0.454 & 0.102 & (0.293, 0.704) & --- \\
\quad Family-level: intercept--slope corr. & $-$0.435 & 0.303 & ($-$0.833, 0.261) & 0.152 \\
\quad Subject-level: intercept & 0.569 & 0.174 & (0.313, 1.036) & --- \\
\midrule
\multicolumn{5}{l}{\textit{Goodness-of-fit (\mlmgof)}} \\
\quad Groups used ($G$) & \multicolumn{4}{l}{5} \\
\quad Wald $\chi^2$ ($df=4$) & 2.922 & & & 0.404 \\
\bottomrule
\end{tabular}
\begin{tablenotes}\small
\item[a] Baseline odds conditional on zero random effects. OR = odds ratio; SD =
  standard deviation; 95\% CIs for random effects are likelihood-ratio-based.
  $G=\min(10,\nmin)=\min(10,5)=5$.
\end{tablenotes}
\end{threeparttable}
\end{table}

\clearpage


\end{document}